\documentclass[twocolumn,showpacs]{revtex4}
\usepackage{graphicx}
\usepackage{enumerate}

\newcommand{\etal}{\textit{et al.}}
\newcommand{\PRL}{\textit{Phys. Rev. Lett.}}
\newcommand{\PR}{\textit{Phys. Rev.}}
\newcommand{\jpb}{\textit{J. Phys. B: At. Mol. Opt. Phys.}} %1988 and onwards
\begin{document}
\title{Schemes for loading a Bose\textendash Einstein condensate into a two-dimensional dipole trap}
\author{Yves Colombe, Demascoth Kadio, Maxim Olshanii$^{\dagger}$, Brigitte Mercier, Vincent Lorent and H{\'e}l{\`e}ne Perrin}
\affiliation{Laboratoire de physique des lasers, CNRS-Universit{\'e} Paris 13\\
99 avenue Jean-Baptiste Cl{\'e}ment, F-93430 Villetaneuse\\
$^{\dagger}$Department~of~Physics~and~Astronomy, University~of~Southern~California,\\
Los Angeles, California 90089-0484, USA}
\begin{abstract}
We propose two loading mechanisms of a degenerate Bose gas into a surface trap. This trap relies on the dipole potential produced by two evanescent optical waves far detuned from the atomic resonance, yielding a strongly anisotropic trap with typical frequencies 40~Hz~$\times~65$~Hz~$\times~30$~kHz. We present numerical simulations based on the time-dependent Gross\textendash Pitaevskii equation of the transfer process from a conventional magnetic trap into the surface trap. We show that, despite a large discrepancy between the oscillation frequencies along one direction in the initial and final traps, a loading time of a few tens of milliseconds would lead to an adiabatic transfer. Preliminary experimental results are presented.
\end{abstract}
\maketitle

\section{Introduction}
\label{intro}
Bose\textendash Einstein condensates (BEC) of alkali atoms \cite{BEC} as sources of coherent matter waves are of considerable interest in atom optics and interferometry. The first atom interferometry experiment using a BEC was performed in 1997~\cite{Ketterle97}. Since then, interferometry has been used to probe the condensate coherence length~\cite{Esslinger00}, to give a signature for the Mott insulator transition in an optical lattice~\cite{Esslinger02} and to test restricted geometry effects~\cite{Ertmer01}. This last example is a witness to the increasing interest in BEC in reduced dimensions~\cite{Ketterle01}.

The 2DEG (two-dimensional electron gas) is a very rich system in condensed matter physics, giving rise for example to the fractional quantum Hall effect~\cite{FQHE} and to anyonic quasi-particles~\cite{anyons}. It is also a convenient medium for electronic interferometry~\cite{Buks98}. By analogy, a 2DAG (two-dimensional atomic gas) would give access to a new regime of quantum degeneracy~\cite{Bagnato91}. Most of the theoretical work investigating this field predicts specific phenomena not encountered in the 3D geometry, such as a progressive coherence below a critical temperature and a modification of the mean field interaction~\cite{Petrov01}. The realization of a 2D condensate is also a preliminary step for the production and the manipulation of anyonic quasi-particles~\cite{Cirac01}. Finally integrated atom optics, where the matter waves can be guided in an arbitrary way, represents an important technological challenge.

The most convenient ways to realize a 2D confinement of alkali atoms use either Zeeman interaction~\cite{Hinds98} or ac Stark optical potentials~\cite{Grimm00}. The Zeeman method is based on the current-carrying micro wires technique which has been used with success to produce BEC in quasi-1D geometries~\cite{ReichelNature01,Zimmermann01}. This technique may be adapted to produce an exponentially decaying B field~\cite{Hinds98}. The advantage is the use of a simple device yielding a large energy spacing of the lowest-lying vibrational levels. The main drawback is the heating produced by proximity magnetic fields above the metallic surface, due to thermal fluctuations. Atoms sitting at distances below 1 micrometre will eventually suffer a high rate of scattering~\cite{Henkel01}. To our knowledge, no experimental demonstration of this suggested mechanism has yet been realized.

On the other hand 2D trapping with dipole forces has been performed in standing waves~\cite{Bouchoule02} or with atoms stopped by an evanescent wave and transferred to the nodes or anti-nodes of a far off-resonant standing wave close to a metallic or dielectric surface~\cite{Pfau98,Spreeuw02}. An inhomogeneous trap in one direction is also naturally produced with two evanescent light waves and the resulting Morse-like potential is conservative for sufficiently large frequency detunings. This idea was proposed more than ten years ago by Ovchinnikov \etal~\cite{Ovchinnikov91}. A good trap geometry is not the only issue for the realization of a 2D gas. An efficient loading of this trap is also essential. A loading scheme of a double evanescent wave trap (DEWT) based on a dissipative transfer of cold atoms has been previously proposed by Dalibard and Desbiolles~\cite{Desbiolles96} and extended for larger frequency detunings~\cite{Perrin00}. Recently, a group in Innsbruck succeeded for the first time in loading a DEWT from a dense dipole trap with 20~000~Cs atoms at a temperature as low as 100~nK~\cite{Grimm02}. All those mechanisms deal with non-condensed atoms. Our paper focuses on the loading of a DEWT from a BEC cloud. We address here two types of loading process and compare their advantages. Our approach is based on adiabatic transformations of combined magnetic plus dipole traps up to the final stage of a DEWT.

The paper is organized as follows: section~\ref{principle} describes the principle of the DEWT and gives our proposed parameters; section~\ref{loading} explains the method of loading starting from two different kinds of trap: (i) the usual magnetic Ioffe\textendash Pritchard trap or (ii) the anti-nodes of a moving red-detuned standing wave as a conveyor belt from the magnetic trap to the DEWT. Details are given in this section about the numerical solutions of the time-dependent Gross\textendash Pitaevskii equation. In section~\ref{section_results}, we present the experimental apparatus and give our preliminary results: production of a $^{87}$Rb BEC near a dielectric surface and transportation of a thermal cloud to this surface.

\section{Principle of the double evanescent wave trap (DEWT)}
\label{principle}
Let us first recall briefly the criterion for reaching the 2D regime for atoms confined in a 3D harmonic trap, with trapping frequencies $\omega_x , \omega_y \ll \omega_z$. In the case of a degenerate Bose gas, the chemical potential $\mu_{\mbox{\scriptsize 3D}}$ calculated for the 3D geometry should fulfil the inequality $\mu_{\mbox{\scriptsize 3D}} \ll \hbar \omega_z$. This leads to a constraint on the atom number $N \ll N^{2D}_{\mbox{\scriptsize BEC}}$ where $N^{2D}_{\mbox{\scriptsize BEC}} = \gamma \, \omega_z^{3/2}/(\omega_x \omega_y)$ \cite{Ketterle01}. With the parameters of $^{87}$Rb one gets $\gamma = 800 \sqrt{2 \pi \, \mbox{rad/s}}$ . By contrast, a 2D classical gas is obtained if $k_{\mbox{\scriptsize B}} T \ll \hbar \omega_z$. This makes sense if the transition temperature $k_{\mbox{\scriptsize B}} T_c \simeq \hbar (\omega_x \omega_y \omega_z N)^{1/3}$ also fulfils this inequality. The requirement on the atom number for a classical 2D gas is thus $N \ll N^{2D}_{\mbox{\scriptsize cl}}$ where $N^{2D}_{\mbox{\scriptsize cl}} = \omega_z^2 / (\omega_x \omega_y)$ whatever the atom. $N^{2D}_{\mbox{\scriptsize cl}}$ is less than $N^{2D}_{\mbox{\scriptsize BEC}}$ as soon as $\omega_z$ is less than $2 \pi \times 640$~kHz for $^{87}$Rb, which is the case in most experiments including ours. We will discuss the validity of the 2D regime in our case later in this section.
 
\begin{figure}[t]
\begin{center}
		\includegraphics[width=7.5cm]{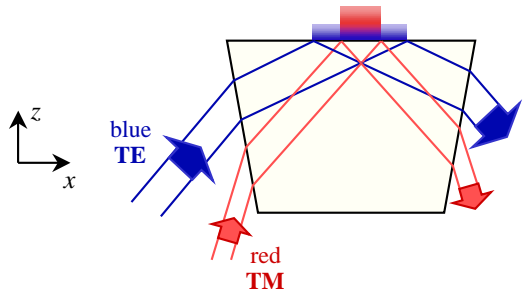}
\caption{Principle of the 2D evanescent light trap. A red-detuned evanescent wave produces a long range attractive exponential potential towards the surface ($z$ axis), and a blue-detuned one prevents atomic adsorption through a short-range repulsive potential. Also shown is our choice of polarizations of the incident beams (TE or TM).}
\label{prism}
\end{center}
\end{figure}

The quasi-2D trap we describe in this paper was first proposed by Ovchinnikov~\etal~\cite{Ovchinnikov91}. It consists in two evanescent light waves produced by total internal reflection at the surface of a dielectric material (see figure~\ref{prism}). One of the light fields is red-detuned by $\delta_r$ with respect to the atomic transition whereas the other one is blue-detuned by $\delta_b$. The angles of incidence of both beams at the dielectric\textendash vacuum interface are chosen such that the decay length of the red field $1/\kappa_r$ is larger than the decay length of the blue field $1/\kappa_b$. The light shift produced by the two fields results in a Morse-like potential along $z$, with a long-range attractive potential and a short-range repulsive wall near the surface. A radial confinement ($x$ and $y$ directions) is achieved with appropriate waists for the red and the blue beams, typically choosing a smaller waist for the red beam. The overall potential seen by the atoms also includes the van der Waals attractive potential towards the dielectric surface.

As our goal is to load such a trap with a degenerate Bose gas, we paid particular attention to keep the spontaneous emission rate as low as possible. The light shift of a field of intensity $I$ \footnote{In the following, we will employ the commonly used term "intensity" instead of the more accurate "irradiance" to refer to power divided by area.}
 detuned by $\delta$ scales as $I/\delta$ whereas the spontaneous scattering rate scales as $I/\delta^2$. At constant light shift, the use of larger detunings and therefore larger intensities is thus preferable. The practical constraints on the availability of laser sources led us to the choice of a YAG laser of a few watts at 1064~nm for the red field and a laser source of a few hundred milliwatts detuned by a few nanometres for the blue field, typically a titanium\textendash sapphire laser. Note that the Innsbruck group came to the same conclusions~\cite{Grimm02}.

We now give the expression of the 2D trapping potential for $^{87}$Rb atoms in the $5S_{1/2}~F=2$ state. The two trapping beams of wavelength $\lambda_r$ and $\lambda_b$ enter a dielectric prism of refraction index $n$ in the $xz$ plane where $z$ is the direction orthogonal to the surface. The angles of incidence at the dielectric\textendash vacuum interface are $\theta_r$ and $\theta_b$, both above the critical angle for total internal reflection $\theta_c = \arcsin(1/n)$. The decay length of the red evanescent wave is then $\kappa_r^{-1} = \sqrt{n^2 \sin^2\theta_r - 1} \; \lambda_r/2 \pi$. A similar expression holds for $\kappa_b^{-1}$.

As the detuning of the YAG laser is large as compared to the fine structure of the excited state $\Delta_{\mbox{\scriptsize FS}}$, we can consider the transition as a $J=0 \longrightarrow J'=1$ transition. The light shift of the ground state due to the red field is always scalar, regardless of the polarization of the evanescent light. Therefore, we choose a TM polarization for this beam, which gives rise to a higher transmission coefficient at the dielectric\textendash vacuum interface.
We will note $\delta_r$ the detuning of the YAG beam with respect to the D2 line at 780~nm. In the following, we do not differentiate between $\delta_r$ and $\delta_r + \Delta_{\mbox{\scriptsize FS}}$.

On the other hand, as the detuning between the blue field and the D2 line is smaller than $\Delta_{\mbox{\scriptsize FS}}$ we have to take into account the contributions of both D1 (at 795~nm) and D2 lines to the light shift. The detuning with respect to the D2 line will be denoted as $\delta_b$, whereas the detuning with respect to the D1 line is $\delta_b + \Delta_{\mbox{\scriptsize FS}}$. As $\delta_b$ is different from $\delta_b + \Delta_{\mbox{\scriptsize FS}}$, the light shift potential will be scalar only if the polarization of the blue evanescent field is linear. In order to ensure a uniform trapping potential for all Zeeman sub-states, we choose a TE polarization for the incoming blue wave, which ensures a linear polarization along $y$ for the evanescent field.

We denote by $P_r$ and $P_b$ the powers of the red and blue beams respectively \textit{inside} the dielectric medium, while $w_r$ and $w_b$ are the beam waists. The corresponding intensities inside the dielectric at the surface are
$I_{r,b} = 2 P_{r,b} \cos \theta_{r,b}/\pi w_{r,b}^2$. The intensity of each beam at $z=0$ in the vacuum depends on its polarization (TE for the blue beam, TM for the red one).

The light shift produced by both beams reads
\begin{eqnarray}
H_{\mbox{\scriptsize LS}}(\mathbf{r}) &=& \hbar \Gamma
\left[ A_r \; e^{\displaystyle -2\kappa_r z} \;
\exp{\left( -2x^2 \cos^2 \theta_r/w_r^2 - 2y^2/w_r^2 \right)} \right. \nonumber \\*
&+& \left. A_b \; e^{\displaystyle -2\kappa_b z} \;
\exp{\left( -2x^2 \cos^2 \theta_b/w_b^2 - 2y^2/w_b^2 \right)} \right]
\label{lightshift}
\end{eqnarray}

where
\begin{eqnarray}
A_r &=& t_{\mbox{\scriptsize TM}}^2 \;\frac{I_r}{2 I_s} \, \frac{\Gamma}{4 \delta_r}\\*
A_b &=& t_{\mbox{\scriptsize TE}}^2 \;\frac{I_b}{2 I_s}
\left( \frac{2}{3} \frac{\Gamma}{4\delta_b}
+ \frac{1}{3} \frac{\Gamma}{4(\delta_b + \Delta_{\mbox{\scriptsize FS}})}\right)
\end{eqnarray}

Note that $A_r <0$. The transmission coefficients for the intensity are
\begin{eqnarray}
t_{\mbox{\scriptsize TM}}^2 &=& \frac{4 n^2 \cos^2 \theta_r}{n^2-1}
\frac{2 n^2 \sin^2 \theta_r -1}{(n^2+1)\sin^2 \theta_r -1}\\*
t_{\mbox{\scriptsize TE}}^2 &=& \frac{4 n^2 \cos^2 \theta_b}{n^2-1}
\end{eqnarray}

Here, $\hbar \Gamma$ is the natural linewidth of the excited state and $I_s = 1.6$~mW/cm$^2$ is the saturation intensity. The terms in $\cos \theta_{r,b}$ in the exponential appear because of the projection of the beam profile onto the dielectric surface.

The total potential includes the van der Waals interaction with the surface. One has to go beyond Lennard-Jones potential in $1/z^3$ because the distance to the surface is comparable to $\lambdabar = \lambda/2\pi$. Therefore retardation effects cannot be ignored. Landragin~\cite{Landragin97} gives an analytical correction to the Lennard-Jones potential which approximates the exact result with a $0.6~\%$ accuracy between 0 and $10~\lambdabar$:
\begin{equation}
H_{\mbox{\scriptsize vdW}}(z) \simeq f(z/\lambdabar) \; H_{\mbox{\scriptsize LJ}}(z)
\end{equation}
with
\begin{eqnarray}
f(u) &=& 0.987 \left( \frac{1}{1+1.098 \, u} - \right. \nonumber \\*
&& \left. \frac{0.00493 \, u}{ 1 + 0.00987 \, u^3 - 0.00064 \, u^4} \right)
\end{eqnarray}
\begin{equation}
H_{\mbox{\scriptsize LJ}}(z) = \frac{n^2-1}{n^2+1} \frac{1}{4 \pi \varepsilon_0} \,
\frac{\langle d^2 \rangle}{12} \frac{1}{\lambdabar} \left(\frac{\lambdabar}{z}\right)^3
\end{equation}

For rubidium in the ground state, the mean value of electric dipole squared is $\langle d^2\rangle = 28.2 \, e^2 a_0^2$ where $a_0$ is the Bohr radius and $e$ the electron charge.

\begin{table}[b]
\caption{\label{values}Values of the 2D trap parameters.}
 \begin{ruledtabular}
 \begin{tabular}{@{}c c c c c c}

	~& $P_{r,b}$ (W) & $w_{r,b}$ ($\mu$m) & $\lambda_{r,b}$ (nm)
	& $\theta_{r,b}$ (deg.) & $\kappa_{r,b}^{-1}$ (nm)\\
 &&&&&\\
	red & 4 & 150 & 1064 & 44.6 & 510 \\
	blue& 0.5 & 170 &  778 & 50 & 220\\

\end{tabular}
\end{ruledtabular}
\end{table}

In the following we give the trap characteristics for a reasonable choice of parameters for the evanescent light, taking 
into account the available power of laser sources near 780~nm. The proposed values of the parameters are indicated in table~\ref{values}. The dielectric medium is chosen to be BK7 glass which has a relatively low index of refraction $n=1.51$ to minimize the van der Waals attraction towards the surface. Using this material, the critical angle for total internal reflection is $\theta_c = 41.5^{\circ}$.

\begin{figure}[t]
\begin{center}
\includegraphics[width=6.2cm]{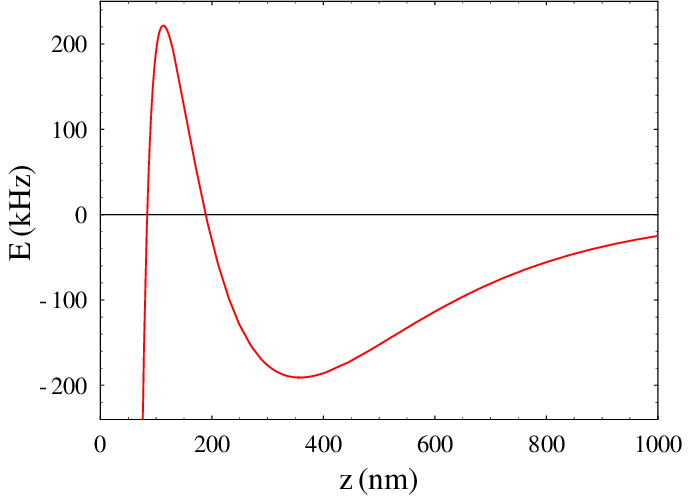}		 \includegraphics[width=6.2cm]{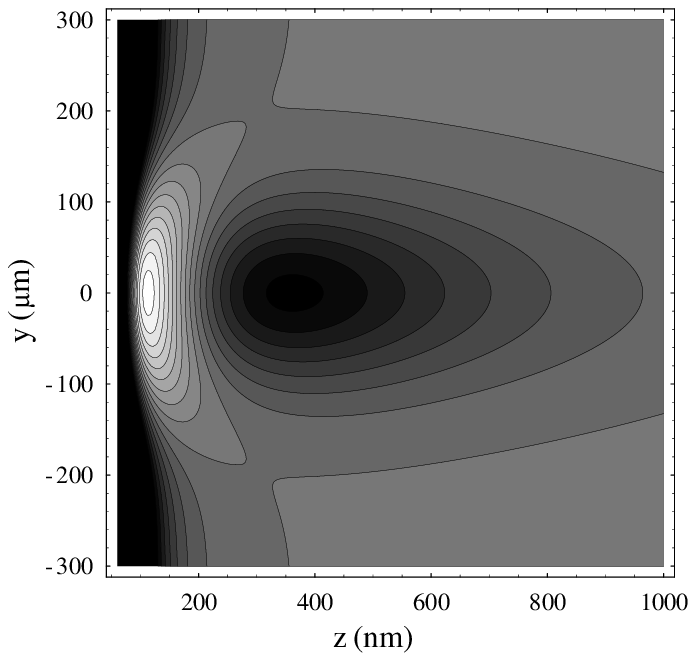}
\caption{2D trapping potential plotted with the values presented in table~\ref{values}. Top: cut along $z$ for $x=y=0$. The red-detuned evanescent wave produces the long-range attractive potential, whereas the blue-detuned evanescent wave acts as a repulsive wall at short distances. The attractive van der Waals potential is dominant very close to the surface ($z<100$~nm). Bottom: contour plot in the $yz$ plane. Subsequent contours are spaced by 25~kHz; darker shades correspond to lower energies.  Horizontal confinment is achieved due to the Gaussian transverse profile of the red incident beam.}
\label{potential}
\end{center}
\end{figure}

These values of the trap parameters lead to a very anisotropic potential above the dielectric surface. A cut of the potential along $z$ for $x=y=0$ is depicted in figure~\ref{potential}, top. The atoms are trapped at a distance $z_0 = 360$~nm from the surface. The trap depth is 180~kHz (or equivalently $9~\mu$K) and is given by the energy difference between the bottom of the trap and the saddle points at $x = 0$, $y = \pm195~\mu$m, $ z = 300$~nm (see the contour plot in figure~\ref{potential}, bottom). We have checked with a 1D numerical calculation that the tunnelling from the ground state to the dielectric surface is negligible. The potential is essentially harmonic around the minimum in the $x$ and $y$ directions where the trapping force results from the transverse profile of both beams. The oscillation frequency is smaller along $x$ due to the angle between the beam axis and the surface. In the $z$ direction the trap deviates rapidly from the harmonic approximation. However, the computed oscillation frequency along $z$ at the bottom of the trap gives a good indication of the anisotropy of the potential: $\omega_x = 2\pi \times 41$~Hz, $\omega_y = 2\pi \times 67$~Hz and $\omega_z = 2\pi \times 28$~kHz. The aspect ratio is thus 690 along $x$ and 420 along $y$. With these parameters, the value of $N^{2D}_{\mbox{\scriptsize BEC}}$ is $1.4 \times 10^6$ and we get either $\mu_{\mbox{\scriptsize 3D}}/h = 10$~kHz for $10^5$~atoms or 4~kHz for $10^4$~atoms, to be compared to 28~kHz. The system is thus already in the 2D regime for a reasonably high number of atoms. However, $N^{2D}_{\mbox{\scriptsize cl}} = 3~\times~10^5$ is lower than $N^{2D}_{\mbox{\scriptsize BEC}}$ and the transition temperature corresponds to about 20~kHz for $10^5$~atoms and 9~kHz for $10^4$~atoms, which means that only thermal clouds close to the transition temperature could be considered as 2D gases.

As this trap is intended to be loaded with a degenerate Bose gas, the spontaneous scattering rate at the bottom of the trap is an important parameter. It is given by the formula
\begin{eqnarray}
&&\Gamma_{\mbox{\scriptsize scatt}} = \Gamma
\left[ t_{\mbox{\scriptsize TM}}^2 \;\frac{I_r}{2 I_s} \, \frac{\Gamma^2}{4 \delta_r^2}
\; e^{\displaystyle -2\kappa_r z_0} \right. \nonumber \\*
&+& \left. t_{\mbox{\scriptsize TE}}^2 \;\frac{I_b}{2 I_s}
\left( \frac{2}{3} \frac{\Gamma^2}{4\delta_b^2}
+ \frac{1}{3} \frac{\Gamma^2}{4(\delta_b + \Delta_{\mbox{\scriptsize FS}})^2}\right) \; e^{\displaystyle -2\kappa_b z_0} \right]
\end{eqnarray}

With our choice of parameters we get $\Gamma_{\mbox{\scriptsize scatt}} = 5$~s$^{-1}$.
This gives a reasonable lifetime for a degenerate gas inside the trap.

\section{Loading the 2D trap}
\label{loading}
The loading of the 2D trap with a degenerate Bose gas is one of the major points
to be addressed in this type of experiment. Methods for loading similar traps with classical gases have been demonstrated before. They rely on optical pumping by evanescent waves in the vicinity of the surface~\cite{Pfau98,Grimm02}. This kind of method is prohibited when dealing with a degenerate gas. In fact, any interaction of the atoms with resonant light should be avoided in order to preserve coherence and reduce heating. In this paper, we propose two different methods for loading the atoms from a magnetic trap, centred a few millimetres above the dielectric surface, into the DEWT. The magnetic trap we start with is a standard cigar-shaped QUIC trap~\cite{Esslinger98} with frequencies $\omega_{\perp} = 2\pi \times 300$~Hz and $\omega_x = 2\pi \times 21$~Hz (see section~\ref{section_exp}). Both schemes rely on the adiabatic transfer from the magnetic trap into the surface trap. We give a brief description of the methods and present results of numerical simulations of the two transfer processes.

\subsection{Scheme 1: magnetic to 2D trap transfer}
\label{scheme1}
The first method consists in deforming a translated magnetic trap adiabatically by switching on the evanescent light trap slowly. The atoms are first translated inside a moving magnetic trap to the vicinity of the surface. This is done by adding to the QUIC trap a pair of Helmholtz coils with a vertical axis (see section~\ref{sec_exp_results}). At the same time, the blue-detuned evanescent light field is switched on in order to prevent the atoms from sticking to the surface as they come very close to it. Second, the condensate is transferred into the 2D dipole trap by switching on the red-detuned evanescent wave slowly, thus compressing the cloud strongly in the $z$ direction. Finally, the magnetic field has to be switched off and the atoms remain trapped in the DEWT. The transfer is adiabatic if the variation of the trap frequency $\omega(t)$ verifies along each direction the following inequality:
\begin{equation}
\frac{\partial \omega}{\partial t} \ll \omega^2
\label{adiabaticity}
\end{equation}
This implies that the ramping time of the evanescent wave must be very large as compared to the oscillating period of the atoms in the trap.

\begin{figure*}[t]
\begin{center}
\includegraphics[width=15cm]{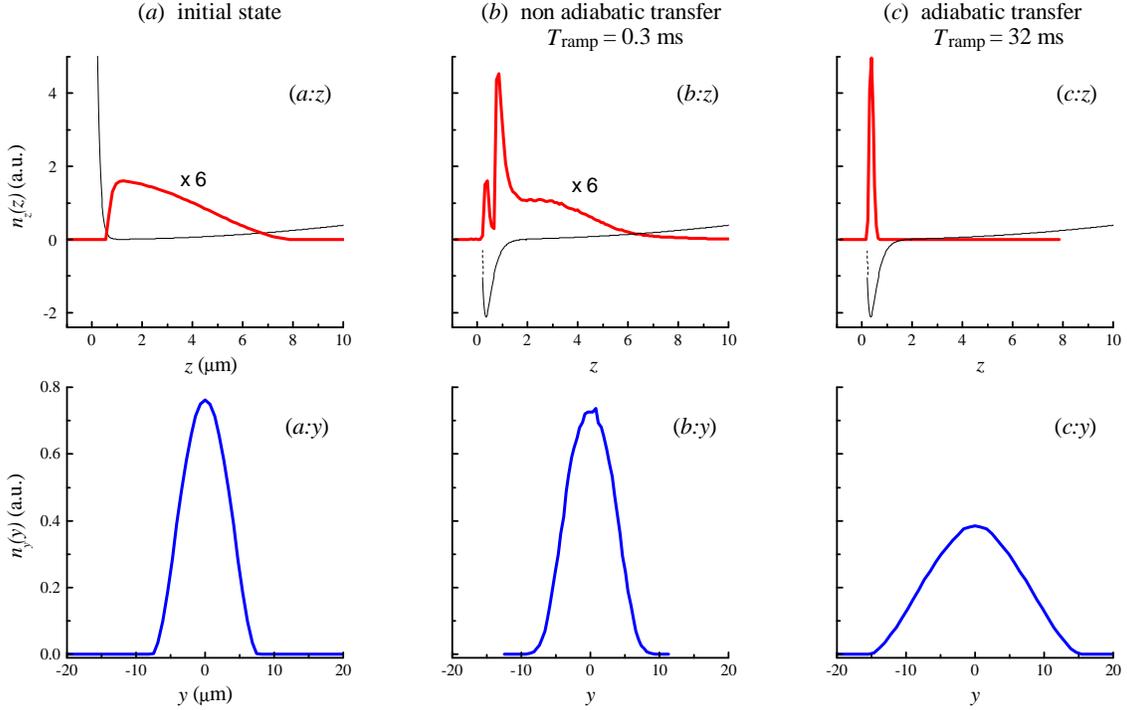}
\caption{Integrated spatial density along $x$ and $y$ as a function of $z$ (({$a$:$z$}) to ({$c$:$z$}), thick curve) and along $x$ and $z$ as a function of $y$ (({$a$:$y$}) to ({$c$:$y$})). Note the scaling factor in figures ({$a$:$z$}) and ({$b$:$z$}). On figures ({$a$:$z$}) to ({$c$:$z$}), the potential is also represented (thin curve). ($a$) shows the initial ground state at $t=0$ with the red beam off. The other parts represent the atomic density after evolution with the GPE for a time $t_{\mbox{\scriptsize end}}$ for two extreme values of $T_{\mbox{\scriptsize ramp}}$: for ($b$) $T_{\mbox{\scriptsize ramp}} = 0.3$~ms and  $t_{\mbox{\scriptsize end}}=2.3$~ms whereas for ($c$) $T_{\mbox{\scriptsize ramp}} = 32$~ms and  $t_{\mbox{\scriptsize end}}=42$~ms. In ($c$:$z$), a Gaussian fit of the density profile gives a $1/e^2$ radius of $\sigma=80$~nm. The radius $R$ obtained by fitting the $y$ profiles with a Thomas\textendash Fermi distribution is $R_a = 7.7~\mu$m, $R_b = 7.9~\mu$m and $R_c = 15.5~\mu$m for ({$a$:$y$}), ({$b$:$y$}), ({$c$:$y$}) respectively.}
\label{sim1results}
\end{center}
\end{figure*}

We present here a 3D numerical calculation of part of this process using the time-dependent Gross\textendash Pitaevskii Equation (GPE) for the macroscopic wavefunction for a condensate comprising $10^6$~atoms. The starting point of the calculation is the ground state of a hybrid trap consisting of the blue-detuned evanescent light plus a magnetic harmonic trap centred at the surface ($z=0$), see figure~\ref{sim1results}($a$). We then let the wavefunction evolve due to the GPE, while switching on the red evanescent field and the van der Waals potential with an exponential time profile. The total ramping time is $T_{\mbox{\scriptsize ramp}}$. The calculation ends at time $t_{\mbox{\scriptsize end}} > T_{\mbox{\scriptsize ramp}}$ and does not include the magnetic field extinction.

We use the splitting method to evaluate the effect of the total Hamiltonian for a time interval $dt$: if $dt$ is small enough, one can let $H_r \, dt$ and $H_p \, dt$ commute, where $H_r$ is the part of the Hamiltonian diagonal in the position basis (potential energy plus interactions) and $H_p$ is diagonal in momentum (kinetic energy). The evolution during $dt$ leads to
\begin{equation}
\psi(t+dt) = T^{\dagger} \exp{\left( \frac{-i H_p \, dt}{\hbar} \right)} T \exp{\left( \frac{-i H_r \, dt}{\hbar} \right)} \psi(t)
\label{schroedinger}
\end{equation}
where $T$ represents a fast Fourier transform and $T^{\dagger}$ its inverse. We calculate the chemical potential $\mu$ and the initial wavefunction by solving the time-dependent GPE with the method of imaginary time; we propagate and renormalize $\psi(\tau)$ at each step using the equation
\begin{equation}
\frac{\partial \psi}{\partial \tau} = - \frac{1}{\hbar} H \psi
\label{imaginary_time}
\end{equation}
to get a positive chemical potential as small as possible. The corresponding wavefunction represents the ground state of the GPE and is taken as initial state $\psi(t=0)$ before deformation. The subsequent evolution is calculated using equation (\ref{schroedinger}) (where $H_r$ is a function of time).

The magnetic trap is chosen to be isotropic with an oscillation frequency $\omega_0 = 2 \pi \times 300$~Hz (oscillation period $T_{\mbox{\scriptsize osc}} = 3.3$~ms) to reduce the calculation time. The 2D trap results from evanescent waves with decay lengths $\kappa_r^{-1} = 510$~nm and $\kappa_b^{-1} = 220$~nm. The light shift at the surface is 5.4~MHz for the blue beam and $-1.4$~MHz for the red one. The oscillation frequencies in the DEWT alone are $\omega_z = 2 \pi \times 30$~kHz along the vertical axis, $\omega_x = 2 \pi \times 30$~Hz and $\omega_y = 2 \pi \times 64$~Hz in the horizontal plane. The position of the minimum of the potential well  along the vertical direction is about 350~nm above the surface of the prism. As $\omega_x$ and $\omega_y$ are much smaller than $\omega_0$, the frequency in the horizontal plane changes only slightly from $\omega_0$ to \textit{e.g.} $\sqrt{\omega_0^2 + \omega_x^2}$ and the criterion for adiabaticity is easily fulfilled horizontally.

The numerical results are shown in figure~\ref{sim1results}. Figures~{\ref{sim1results}\,($a$:$z$)} to {\ref{sim1results}\,($c$:$z$)} represent the spatial atomic density integrated along $x$ and $y$ as a function of $z$ together with the trapping potential along $z$. Figures {\ref{sim1results}\,($d$:$y$)} to {\ref{sim1results}\,($c$:$y$)} show the density integrated along $x$ and $z$ and plotted as a function of $y$. Figure~{\ref{sim1results}\,($a$)} represents the initial atomic distribution ($t=0$) of the $10^6$ atoms, corresponding to the ground state of the hybrid magnetic plus blue evanescent wave trap. The atomic density along $z$ is not symmetrical due to the steep wall produced by the blue beam. The spatial density for two different ramping times $T_{\mbox{\scriptsize ramp}}$ of the red beam is given in figures~{\ref{sim1results}\,($b$)} and {\ref{sim1results}\,($c$)}.

We observe that almost all the condensate is transferred into the dipole trap for a ramping time $T_{\mbox{\scriptsize ramp}} = 32$~ms much greater than the initial oscillation period $T_{\mbox{\scriptsize osc}}$, figure~{\ref{sim1results}\,($c$)}. We also note that, in this case, the width of the spatial density along the horizontal direction is enlarged when adding the red beam, compare figure~{\ref{sim1results}\,($c$:$y$)} with figure~{\ref{sim1results}\,($a$:$y$)}. Indeed when ramping up the red beam intensity slowly, the atomic cloud is compressed along the vertical axis, figure~{\ref{sim1results}\,($c$:$z$)}, and expands in the horizontal directions where the potential energy increases more slowly with distance from the centre.

\begin{figure}[t]
\begin{center}
\includegraphics[width=7.5cm]{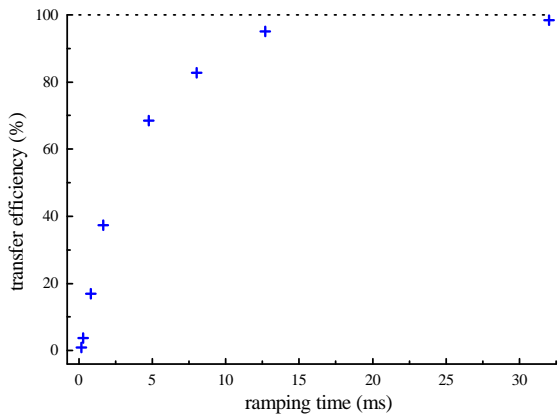}
\caption{Transfer efficiency as a function of the ramping time. The crosses represent numerical results. The dotted line represents 100~\% efficiency.}
\label{transfer}
\end{center}
\end{figure}

On the other hand, for a very short ramping time (0.3~ms), only a few atoms are transferred into the 2D dipole trap (figure~{\ref{sim1results}\,($b$})). The integrated atomic density along $z$ presents three features: we observe two density peaks with a large difference between their amplitude, plus a broad background. The centre of the smaller peak corresponds to the minimum of the dipole trap and represents the atoms which were transferred successfully. The larger peak sits at the extreme border of the dipole trap well. One can understand this with a simple picture. When we ramp up the red beam quickly, the atoms which were localized at the bottom of the initial potential with almost zero velocity suddenly acquire an energy equal to the depth of the dipole potential. They populate a very narrow band of excited states of the new well and are essentially located at the turning points of the potential. Towards the surface, the potential is very stiff and the atomic velocity changes almost instantaneously. As a result, a single density peak is visible at the right turning point where the potential is much shallower. Note that the respective weight of the two peaks changes as the loading time $T_{\mbox{\scriptsize ramp}}$ is increased: more atoms are transferred in the DEWT, see figure~\ref{transfer}, and the second peak is reduced. The background of the distribution represents the atoms which are not much affected by the switching on of the red beam. These atoms were out of the well in the right-hand side of the potential when the red beam was ramped up. The width of the spatial density along the horizontal direction (figure~{\ref{sim1results}\,($b$:$y$})) does not change as compared to figure~{\ref{sim1results}\,($a$:$y$)} due to the poor compression reached in the $z$ direction.

The analysis of the transfer efficiency is presented in figure~\ref{transfer}. To estimate the number of transferred atoms, we fit the peak in the $z$ density profile corresponding to the atoms trapped in the DEWT with a Gaussian profile and compare its area to the total number of atoms. The transfer efficiency increases non linearly with the ramping time. For a ramping time $T_{\mbox{\scriptsize ramp}}= 32$~ms, very large as compared to $T_{\mbox{\scriptsize osc}}$, the difference between the final density profiles and the ground state of the total trap (DEWT + magnetic trap) is imperceptible and the transfer is actually adiabatic. From figure~\ref{transfer} we infer that a 20~ms ramping time is quite enough to realize an adiabatic transfer. Note that in the adiabatic case, as we ramp the magnetic trap off very slowly, the spatial density along the $z$ axis becomes narrower while the horizontal width becomes larger (not shown here). Indeed, the horizontal frequencies of the dipole trap ($\omega_x = 2\pi \times 30$~Hz, $\omega_y = 2\pi \times 64$~Hz) are smaller than those in the initial trap ($\omega_x = \omega_y = 2\pi \times 300$~Hz), and the anisotropy is more pronounced due to the strong confinement along $z$ and the interactions between atoms.

\subsection{Scheme 2: the atomic lift}
\label{scheme2}
\begin{figure*}[t]
\begin{center}
\vspace{0.5cm}	
\includegraphics[width=15cm]{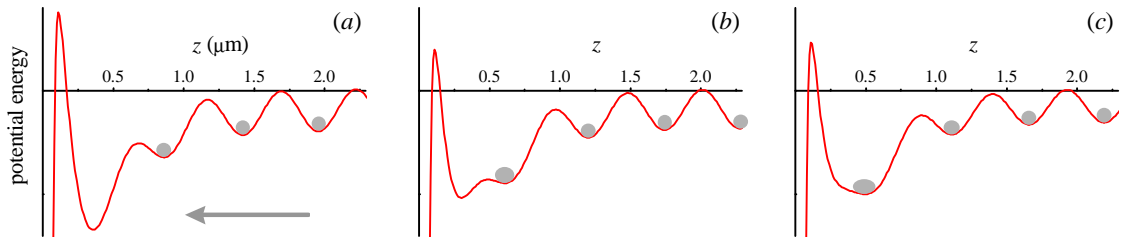}
\caption{Principle of the atomic lift. Three different stages are shown for three values of the relative phase $\varphi$ between the two beams of the standing wave. ($a$) $\varphi = 0.6 \pi$; ($b$) $\varphi = 1.5 \pi$; ($c$) $\varphi = 1.8 \pi$.}
\label{lift}
\end{center}
\end{figure*}

The second method makes use of a moving standing wave to transport the atoms from the magnetic trap into the DEWT. The idea is close to the principle of the conveyor belt recently realized on a micro chip~\cite{Reichel01}. At the beginning of the loading process, the atoms are confined in the magnetic trap and the laser fields producing the DEWT are on.
The process may be decomposed into three steps: (i) The atoms are loaded at the anti-nodes of a stationary wave into a series of horizontal planes obtained by two red-detuned counter-propagating beams along $z$.
(ii) The magnetic field is switched off and the atoms are lifted down towards the surface by changing the phase of one of the beams with respect to the other. The atoms accumulate in the 2D trap by continuous deformation of the potential (figure~\ref{lift}). (iii) Finally, when all the atoms are in the last well near the surface, the stationary wave is switched off and the atoms remain trapped in the DEWT.

At each step, the different transfers have to remain adiabatic to avoid any heating of the atomic cloud. Step (i) should not pose particular problems as transfer of condensates into optical lattices has already been demonstrated~\cite{Inguscio01}. Note that the atoms remain trapped horizontally even after the magnetic field has been switched off due to the Gaussian profile of the counter-propagating beams, with corresponding horizontal oscillation frequency $\omega_h$. One simply has to switch off the magnetic field adiabatically, that is in a time larger than $\omega_h^{-1}$.

\begin{figure*}[t]
\begin{center}
\vspace{0.5cm}		\includegraphics[width=15cm]{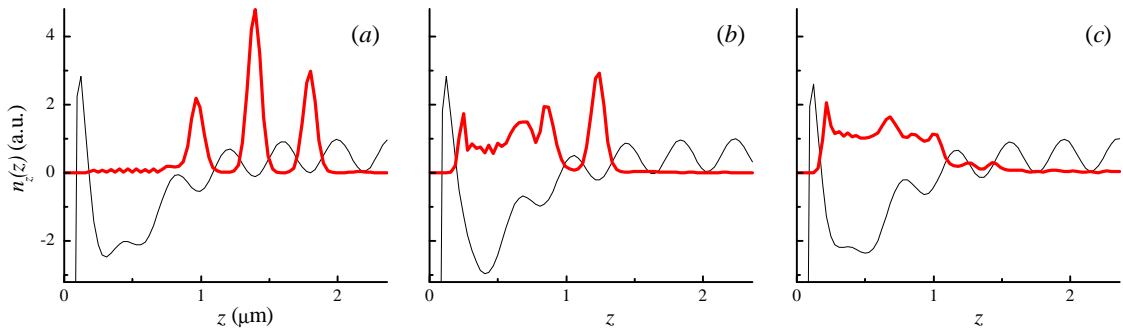}
\caption{Results of the numerical simulation of the atomic lift: integrated spatial density along $x$ and $y$ as a function of $z$ (thick curve), at three evolution times: ($a$) $t=5.8$~ms; ($b$) $t=13.3$~ms; ($c$) $t=22.3$~ms. The potential is also represented (thin curve).}
\label{lift_results}
\end{center}
\end{figure*}

To realize this first step, one can typically use a $\lambda_{\mbox{\scriptsize lift}}=830$~nm laser diode
with a power of 15~mW in each beam and a waist of $90~\mu$m. This gives $\omega_h=2\pi \times 60$~Hz. Provided the populated anti-nodes are far from the evanescent waves, the phase change may be very fast, and this gives the atoms a large translational velocity, $v_1$. When the atoms approach the surface, the velocity must be lowered, to $v_2$ say, because the horizontal shape of the trap changes. In fact the horizontal frequency seen by an atom changes, in the $x$ direction for example, from $\omega_h$ to $\omega_{\mbox{\scriptsize max}} = (\omega_h^2 + \omega_x^2)^{1/2}$ when it is first loaded in the DEWT, and then changes periodically between $\omega_{\mbox{\scriptsize max}}$ and $\omega_x$ as the phase evolves further to load the remaining atoms. This gives the typical time constant for adiabaticity.

We tested this loading scheme qualitatively with a 3D numerical simulation analogous to the one described in the last section. We were interested mostly in the last stage where the atomic cloud distributed in several planes approaches the surface. The initial state of the calculation ($t=0$) is the ground state of the GPE in the potential formed by the stationary wave with $\lambda_{\mbox{\scriptsize lift}} = 800$~nm plus a 1D harmonic potential along $z$. This potential mimics the situation immediately after the transfer from the magnetic trap into the stationary wave. The 1D potential is centred $3.4~\mu$m above the surface and is switched off at the beginning of the evolution through the time-dependent GPE, while the 2D trap is switched on. At this point ($t=1.1$~ms), the atoms are spread over a few planes. For an initial vertical oscillation frequency of 300~Hz and $10^5$ atoms, typically 15 planes are populated. However, to reduce the calculation time, we start from a condensate in a harmonic trap with a vertical oscillation frequency of 1.6~kHz. In this case, the atoms are spread symmetrically over three planes (see figure~\ref{lift_results}). The relative phase between the two beams of the standing wave is then allowed to evolve, resulting in the two successive velocities $v_1=1.1$~mm/s and $v_2=76~\mu$m/s, as mentioned above.

The results of the numerical calculation are shown in figure~\ref{lift_results}. The column density integrated over $x$ and $y$ is plotted as a function of $z$ at three stages of the loading process. At $t=5.8$~ms ($a$) the first populated plane reaches the rim of the last well, which coincides with the 2D trap. At $t=13.3$~ms ($b$) two populated planes have melted into the last well. At $t=22.3$~ms ($c$) the atoms accumulate in the last well. However, it was not possible to fulfil the condition for adiabatic transfer. As a result, the final atomic density does not coincide with the ground state of the GPE in the DEWT. We shall discuss this point further in the conclusion. Moreover, when the atoms are in the last well of the standing wave before melting, they tunnel through the small barrier separating them from the trap, as can be seen already in ($a$), thus populating excited states of the DEWT. To limit this phenomenon, both the depth and the typical size of the wells of the standing wave must be adjusted to those of the DEWT.
This condition may be a difficult point to address experimentally.

\section{Preliminary experimental results}
\label{section_results}
In this section, we present some preliminary experimental results towards loading
of a degenerate Bose gas into a DEWT.
\subsection{Experimental set-up}
\label{section_exp}
\begin{figure}[t]
\begin{center}
		\includegraphics[width=5cm]{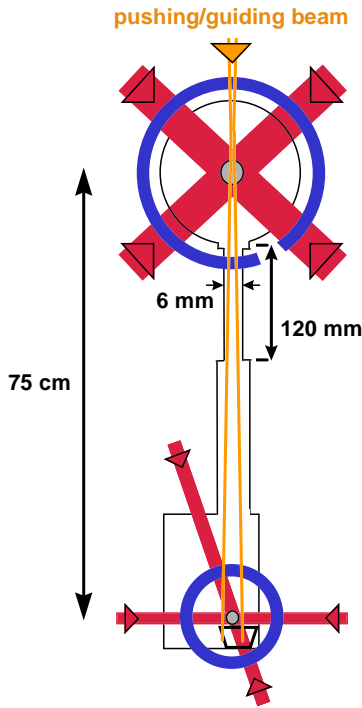}
\caption{Schematic view of the experimental set-up.}
\label{setup}
\end{center}
\end{figure}

The experimental set-up, figure~\ref{setup}, consists of two ultra-high vacuum chambers separated vertically by 75~cm. In the upper chamber (pressure $10^{-9}$~Torr) the $^{87}$Rb atoms are collected from a vapour in a standard magneto-optical trap (MOT) by three retro-reflected beams ($1/e^2$-diameter 25~mm each, total power 45~mW). This MOT acts as a reservoir for the second MOT in the lower cell where the pressure is below $10^{-11}$~Torr. The difference in vacuum pressure is ensured by a tube of length 120~mm with an internal diameter of only 6~mm. The lower MOT is set up with six independent laser beams with a $1/e^2$-diameter of 10~mm each for a total power of 45~mW.\\

The atoms are continuously transferred into the second cell by an original method combining pushing and guiding the atoms. The pushing beam has a power of 30~mW and is detuned by 2.5~GHz on the red side of the MOT $5S_{1/2},~F=2 \longrightarrow 5P_{3/2},~F'=3$ transition. It is focused 8~cm above the upper MOT to a waist of $220~\mu$m such that its radius is $250~\mu$m at the upper MOT and $940~\mu$m at the lower MOT. This beam induces sufficiently large light shifts (30~MHz) so that atoms inside the beam no longer feel the MOT beams; the atoms are extracted from the upper MOT with a radiation pressure about 25 times smaller than that of a typical MOT. The advantage of this method is that the velocity of atoms remains in the capture range of the lower MOT (about 15~m/s). Furthermore, the pushing beam acts as a dipole trap which guides the atoms vertically near its axis inside the small diameter tube. The depth of this guide is about 1.4~mK at the upper MOT for the $F=2$ state. A collinear repumping beam, tuned to the $F=1 \longrightarrow F'=2$ transition, ensures that the atoms remain in $F=2$. Due to the divergence of the guiding beam the radiation pressure at the position of the lower trap is negligible. The fact that the atoms are guided while pushed towards the lower cell renders the loading process very robust against small changes in the parameters of the two MOTs.

After 30~s of loading time, the atoms are cooled by molasses cooling, compressed and pumped into the $F=2, m_F=2$ state. They are then transferred into a Ioffe\textendash Pritchard, cigar-shaped magnetic trap. The coils producing the magnetic field are placed following the quadrupole and Ioffe configuration (QUIC)~\cite{Esslinger98} and dissipate a total electric power of 200~W. They are cooled by thermal contact with water-cooled copper radiators. Two independent current sources are used, one for the Ioffe coil (43~A) and one for the quadrupole coils (30~A). The resulting magnetic field has a minimum $B_0$ of about 1~G with a field gradient $b'=225$~G/cm and a curvature $b''=270$~G/cm$^2$. The resulting oscillation frequencies are 21~Hz along the axis of the cigar ($x$ direction) and 300~Hz in the transverse $yz$ plane for the $F=2, m_F=2$ state. We achieve Bose\textendash Einstein condensation after 30~s of RF evaporative cooling.

\subsection{Results}
\label{sec_exp_results}
\begin{figure}[t]
\begin{center}
\includegraphics[width=7.5cm]{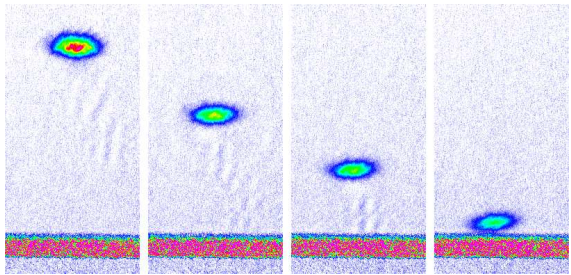}
\caption{Absorption images of a thermal cloud at temperature 9~$\mu$K approaching the prism. Each 2.5~mm~$\times$~5~mm image is taken 4~ms after switching off the magnetic fields, at different steps of a 1~s, 0 to 20~A current ramp in the additional coils. The final values of the current are 0, 10, 15 and 18.4~A respectively. When reaching 20~A, the atoms are lost by contact with the surface in the absence of a blue-detuned evanescent wave.}
\label{exp_results}
\end{center}
\end{figure}

The condensate contains typically $2 \times 10^5$ atoms and is produced at the centre of the QUIC trap, just above the prism which is the support for the 2D evanescent light trap (see the bottom of figure~\ref{setup}). The vertical position of the prism with respect to the condensate may be adjusted mechanically. However, it is limited to at least 3~mm to maintain a correct loading efficiency of the MOT. By adding to the QUIC trap a uniform vertical magnetic field varying from 0 to 64~G the minimum of the trapping potential is moved down to the surface of the prism, as required by the first 2D-transfer scheme. This field is produced by ramping up a current from 0 to 20~A in two additional horizontal coils separated by 4~cm, having each 30 windings of mean diameter 16~cm. During the ramp, the current in the quadrupole coils of the QUIC trap is lowered by about 5\%; otherwise the trap would cross a zero of the magnetic field and the atoms would separate into two clouds. The resulting potential near the centre of the trap remains essentially unaltered apart from an increase of a factor of 1.6 in the oscillation frequency along the slow axis (from 21~Hz to 33~Hz). This axis is also slightly tilted, in agreement with the magnetic field calculations. To illustrate the translation process, we filled the initial QUIC trap with a sample of thermal atoms at $T=9~\mu$K centered 3.7~mm above the prism and imaged the atomic cloud at several steps of its journey towards the surface, figure~\ref{exp_results}. When the current in the additional coils reaches 20~A, the atoms are lost by contact with the surface in the absence of a blue-detuned evanescent wave.

As the experiment was performed with a non-condensed cloud, the next step will be to extend it to a BEC. The blue and red beams have then to be added to the experimental set-up. Preliminary experiments with a blue beam alone were realized: a bounce of thermal atoms released from the QUIC trap 3.7~mm above the prism was observed. However, the detuning $\delta_b = 2 \pi \times 500$~MHz was too small for the beam to be used in a loading experiment and it will be replaced by an appropriate laser source. This should allow us to test experimentally the first transfer scheme.

\section{Conclusion}
\label{conclusion}
In this paper, we studied two possible methods for loading a degenerate Bose gas into a strongly anisotropic trap. The first method is also the simpler one. We have demonstrated experimentally with a thermal cloud the first stage corresponding to the vertical translation of the atomic cloud. We simulated the transfer from the translated QUIC trap into the DEWT. When the criterion for adiabaticity is fulfilled the transfer efficiency is close to 100\%. The simulation assumed an isotropic magnetic trap with oscillation frequency $\omega_0 = 2 \pi \times 300$~Hz. In the experimental situation, one of the horizontal frequencies is $\omega_{0x}= 2 \pi \times 33$~Hz only and the loading time may have to be increased by a small factor. The last stage corresponds to the extinction of the magnetic field and should not pose particular problems providing that the switching time is much longer than the horizontal frequencies in the DEWT. We believe that this loading method could be implemented experimentally. The difficulty is to adjust the horizontal position of the magnetic trap to the center of the DEWT. This can be done by changing slightly the balance between the currents in the QUIC coils.

The second method has the great advantage of compressing the atomic cloud already in the QUIC trap before the atoms are loaded into the DEWT. The translation stage towards the surface (velocity $v_1$) may be faster than in the last method, as the vertical oscillation frequency in the series of planes is on the order of a few tens of kHz instead of 300~Hz. The accumulation stage into the DEWT by a phase sweep seems to be an elegant idea. However, the phase velocity $v_2$ is strongly limited if one tries to fulfil the adiabaticity condition. In fact, for horizontal frequencies $\omega_h = 2 \pi \times 60$~Hz in the standing wave and $\omega_x = 2 \pi \times 30$~Hz in the DEWT, the frequency change between $\omega_x$ and $\sqrt{\omega_h^2 + \omega_x^2}$ has to be slower than $\omega_x$. Therefore the velocity $v_2$ has to be much smaller than $\lambda_{\mbox{\scriptsize lift}}/4 \times \omega_x/2 \pi = 6~\mu$m/s to avoid 2D breathing mode excitation. The typical time to load 15 planes is then a few seconds. This is still an order of magnitude too large to be of interest for practical applications. To reduce this time, one may either increase the horizontal frequency during the loading process, or compress the cloud initially to populate fewer horizontal planes. Another problem that will occur with this loading method is the phase noise in the standing wave. The study of its influence on the loading mechanism is out of the scope of this paper. However, a recent work in Bonn showed that it produces heating and reduces the lifetime in a standing wave trap~\cite{Alt03}.

Finally, let us note that the second method may be used in a reversed way to extract the 2DAG and study it far from the surface: once the atoms are confined in the DEWT, one can slowly switch on a standing wave and change the relative phase to lift the atoms away from the surface. It is then easier to produce a single 2D trap with a very high aspect ratio with less laser power.

\acknowledgments{We are indebted to Y.~Castin and D.~Lemoine for fruitful discussions, and to R.~J.~Butcher for a critical reading of the manuscript. We acknowledge support from the R{\'e}gion Ile-de-France (contract number E1213). Laboratoire de physique des lasers is UMR~7538 of CNRS and Paris 13 University.}

\end{document}